\documentclass[nofootinbib,twocolumn]{revtex4}
\usepackage{graphicx,amsmath,slashed}

\newcommand{\beq}{\begin{eqnarray}}
\newcommand{\eeq}{\end{eqnarray}}
\newcommand{\yut}{(Y_u Y_u^\dagger)_{\slashed{\mathrm{tr}}}}
\newcommand{\ydt}{(Y_d Y_d^\dagger)_{\slashed{\mathrm{tr}}}}
\newcommand{\au}{\mathcal{A}_u}
\newcommand{\ad}{\mathcal{A}_d}
\newcommand{\had}{\mathcal{\hat A}_d}
\newcommand{\hadp}{\mathcal{\hat A}'_d}

\newcommand{\aud}{\mathcal{A}_{u,d}}
\newcommand{\haud}{\mathcal{\hat A}_{u,d}}
\newcommand{\haudp}{\mathcal{\hat A}'_{u,d}}
\newcommand{\hD}{\mathcal{\hat {\vec D}}}
\newcommand{\tr}{\mathrm{tr}}
\newcommand{\hj}{\hat J}
\newcommand{\hjd}{\hat J_d}

\newcommand{\hjud}{{\hat J}_{u,d}}
\newcommand{\hjq}{\hat J_Q}
\newcommand{\ltev}{\left( \frac{\Lambda_{\rm NP}}{1\,\textrm{TeV}} \right)}
\newcommand{\xqo}{X_Q^{\Delta F=1}}
\newcommand{\xqt}{X_Q^{\Delta F=2}}
\newcommand{\dmqot}{\Delta\tilde m^2_{Q_2Q_1}}
\newcommand{\dmqoth}{\Delta\tilde m^2_{Q_3Q_1}}
\newcommand{\dmqij}{\Delta\tilde m^2_{Q_iQ_j}}

\begin{document}

\title{\bf Covariant Description of Flavor Violation at the LHC}

\author{Oram Gedalia}
\affiliation{Department of Particle Physics and Astrophysics,
Weizmann Institute of Science, Rehovot 76100, Israel}
\author{Lorenzo Mannelli}
\affiliation{Department of Particle Physics and Astrophysics,
Weizmann Institute of Science, Rehovot 76100, Israel}
\author{Gilad Perez}
\affiliation{Department of Particle Physics and Astrophysics,
Weizmann Institute of Science, Rehovot 76100, Israel}
\begin{center}
\begin{abstract}
\centerline{{\bf Abstract}} A simple formalism to describe
flavor and CP violation in a model independent way is provided.
Our method is particularly useful to derive robust bounds on
models with arbitrary mechanisms of alignment. Known
constraints on flavor violation in the $K$ and $D$ systems are
reproduced in a straightforward and covariant manner.
Assumptions-free limits, based on top flavor violation at the
LHC, are then obtained. In the absence of signal, with
100\,fb$^{-1}$ of data, the LHC will exclude weakly coupled
(strongly coupled) new physics up to a scale of 0.6\,TeV
(7.6\,TeV), while at present no general constraint can be set
related to $\Delta t=1$ processes. $\Delta F=2$ contributions
will be constrained via same-sign tops signal, with a model
independent exclusion region of 0.08\,TeV (1.0\,TeV). However,
in this case, stronger bounds are found from the study of CP
violation in $D-\overline D$ mixing with a scale of 0.57\,TeV
(7.2\,TeV). We also apply our analysis to supersymmetric and
warped extra dimension models.

\end{abstract}
\end{center}
\maketitle

\section{Introduction}
The standard model (SM) has a unique way of incorporating CP
violation (CPV) and suppressing flavor changing neutral
currents (FCNCs). Till today no deviation from the SM
predictions related to quark flavor violation has been
observed. Regarding the first two generations, models which do
not include some sort of degeneracies or flavor alignment (that
is, when new physics contributions are diagonal in the quark
mass basis) are bounded to a high energy scale. Moreover,
contributions involving only quark doublets cannot be
simultaneously aligned with both the down and the up mass
bases, hence even alignment theories are constrained by
measurements. However, the hierarchy problem is not triggered
by the light quarks, but rather by the large top Yukawa, where
almost any natural new physics (NP) model consists of an
extended top sector. Ironically, the top flavor sector is the
least understood one, and at present no model independent bound
on its coupling is known to exist.

In this work, we formulate a simple basis independent formalism
for studying flavor constraints in the quark sector (recent
related work about algebraic flavor invariants can be found
in~\cite{Feldmann:2009dc,Jenkins:2009dy}). We start with a two
generations analysis, where a natural geometric interpretation
can be applied. It allows us to straightforwardly reproduce
known results~\cite{Blum:2009sk}. We then consider the three
generations case, where a dramatic improvement in the
measurements related to the top sector is expected at the LHC.
The combination of data from the down and the up sectors is
used to robustly constrain models including arbitrary
mechanisms of alignment.

In the absence of Yukawa interactions, the SM quark sector
possesses a global $G_{\rm SM}=U(3)_Q\times U(3)_U \times
U(3)_D$ flavor symmetry, where $Q$, $U$ and $D$ stand for quark
doublets, up and down type quark singlets, respectively.
$G_{\rm SM}$ is broken by the Yukawa couplings $Y_u$ and $Y_d$,
which transform as $(\mathbf{3}, \bar{\mathbf{3}},1)$ and
$(\mathbf{3},1,\bar{\mathbf{3}})$, respectively, under the
flavor group. The spurions $Y_u Y_u^\dagger$ and $Y_d
Y_d^\dagger$ are then both in the $(\mathbf{8}+1,1,1)$
representation. Since the trace of these matrices does not
affect flavor changing processes, it is useful to remove it,
and work with $\yut$ and $\ydt$, adjoints of $SU(3)_Q$. For
simplicity of notation, we denote these objects as
\beq
\au \equiv \yut \, , \qquad \ad \equiv \ydt \, .
\eeq

\section{Two Generations}
Any hermitian traceless $2\times 2$ matrix can be expressed as
a linear combination of the Pauli matrices. This combination
can be naturally interpreted as a vector in 3D real space,
which applies to $\ad$ and $\au$. We can then define a length
of such a vector, a scalar product, a cross product and an
angle between two vectors, all of which are basis independent:
\beq \label{definitions}
\hspace*{-.3cm}&&|\vec{A}|^2 \equiv \frac{1}{2} \tr(A^2),  \
\vec{A} \cdot \vec{B} \equiv \frac{1}{2} \tr(A  B), \  \vec{A}
\times \vec{B} \equiv \frac{i}{2} \left[ B,A \right] , \nonumber \\
\hspace*{-.3cm}&&\cos \theta_{AB} \equiv \frac{\vec{A} \cdot
\vec{B}}{|\vec{A}| |\vec{B}|}, \ \sin \theta_{AB} =\frac{\left|
\vec{A} \times \vec{B} \right|}{|\vec{A}||\vec{B}|}\, ,
\eeq
where the two angle definitions are equivalent. This allows for
an intuitive understanding of the flavor and CPV induced by a
NP source. Consider a dimension six $SU(2)_L$-invariant
operator, involving only quark doublets,
\beq \label{o1}
\frac{z_1}{\Lambda_{\rm NP}^2}\, O_1=\frac{1}{\Lambda_{\rm
NP}^2} \left[ \overline{Q}_{i} (X_Q)_{ij} \gamma_\mu Q_{j}
\right] \left[ \overline{Q}_{i} (X_Q)_{ij} \gamma^\mu Q_{j}
\right] \, ,
\eeq
where $\Lambda_{\rm NP}$ is some high energy scale and $z_1$ is
the Wilson coefficient. $X_Q$ is a traceless hermitian matrix,
transforming as an adjoint of $SU(3)_Q$ (or $SU(2)_Q$ for two
generations).

\begin{figure}[htb]
\centering
\includegraphics[width=2.33In]{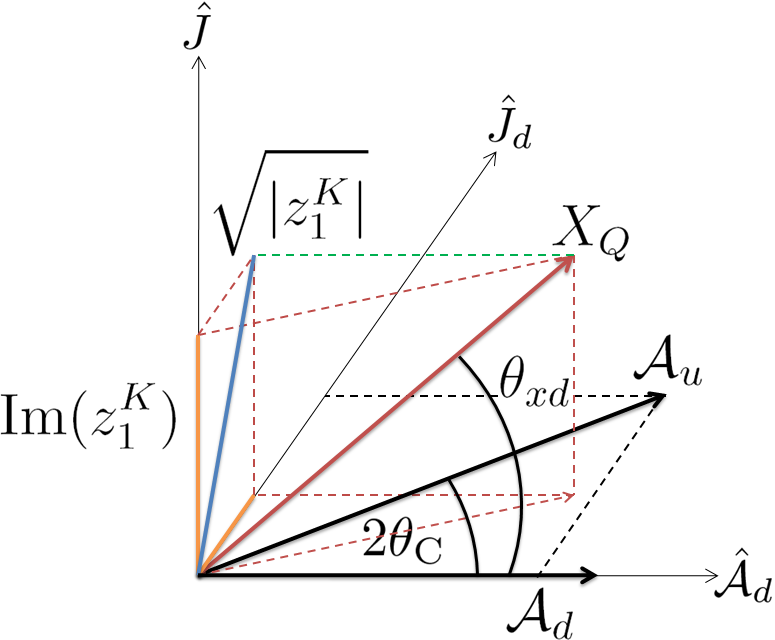}
\caption{Flavor violation in the Kaon system induced by $X_Q$.
The overall contribution to $K^0-\overline{K^0}$ mixing is given by
the solid blue line. The CPV contribution, $\mathrm{Im}(z_1^K)$, is twice
the product of the two solid orange lines, which are the projections
of $X_Q$ on the $\hj$ and $\hjd$ axes. Note that the angle between $\ad$
and $\au$ is twice the Cabibbo angle, $\theta_{\rm C}$.}
\label{fig:2g}
\end{figure}

The contribution to $\Delta c,s\,=\,2$ transitions due to $X_Q$
is given by the misalignment between it and $\aud$, and it is
easy to see that this is equal to
 \beq \label{2g_fv}
\left| z_1^{D,K} \right|=  \left| { X_Q} \times {\aud}
\right|^2/ |{\aud}|^2 = \left| { X_Q} \times {\haud} \right|^2
\,,
\eeq
where ${\haud} \equiv \aud /\big|\aud \big| \,.$ This result is
manifestly invariant under a change of basis. Next we move to
CPV
\beq \label{imz1k_def}
\mathrm{Im}\left(z_1^{K,D}\right)=2\left(X_Q \cdot \hat
J\right)\left( X_Q \cdot \hat J_{u,d} \right)\,,
\eeq
where $\hat J\equiv \ad \times \au/\left|\ad \times \au\right|$
and $ \hat J_{u,d} \equiv  \haud\times \hat J/\left|
\haud\times \hat J \right|$. The above spurions and observables
are easily described geometrically, say in the $\had-\hat
J-\hat J_d$ space, as shown in Fig.~\ref{fig:2g}.
To derive the weakest bound, we express $X_Q$ in terms of its
components
\beq\label{XQ}
X_Q=X^{u,d} \haud+X^J \hj+X^{J_{u,d}} \hjud \,,
\eeq
where we have $X^u=\cos 2\theta_{\rm C} X^d-\sin 2\theta_{\rm
C} X^{J_d}$, $X^{J_u}=-\sin 2\theta_{\rm C} X^d-\cos
2\theta_{\rm C}  X^{J_d}$ and $X^J$ remains invariant. Plugging
the expression for $X_Q$ from Eq.~\eqref{XQ} into
Eqs.~\eqref{2g_fv} and~\eqref{imz1k_def}, one easily reproduces
the results of~\cite{Blum:2009sk} derived in a specific basis.

A new condition for CPV is implied, exclusively related to
$\Delta c,s\,=\,2$ processes and not to $\Delta c,s\,=\,1$
ones:
\beq \label{cpv_cond}
X^{J_{u,d}} \propto \tr \left(X_Q [\aud,[\ad,\au]] \right)
\neq 0 \,,
\eeq
while $X^J\neq0$ provides a necessary condition for all types
of two generations CPV~\cite{Blum:2009sk}. The conditions are
physically transparent and involve only observables, where the
weakest bound on NP is derived for the ratio $X^d/X^{J_d}$
given a fixed amount of CPV, $X^J$. Note, however, that this
new condition in Eq.~\eqref{cpv_cond} is only applicable to
either the down or the up sector, while $X^J\neq0$ is
universal.

\section{Three Generations}
For three generations, a simple 3D geometric interpretation
does not naturally emerge anymore, as the relevant space is
characterized by the eight Gell-Mann matrices. A useful
approximation appropriate for third generation flavor violation
is to neglect the masses of the first two generation quarks,
where the breaking of the flavor symmetry is characterized by
$[U(3)/U(2)]^2$~\cite{Kagan:2009bn}. It is especially suitable
for the LHC, where it would be difficult to distinguish between
light quark jets of different flavor. In this limit the CKM
matrix is reduced to a real matrix with a single rotation angle
between an active light flavor (say, the 2nd one) and the 3rd
generation,
\beq
\theta\cong \sqrt{\theta_{13}^2+\theta_{23}^2}\,,
\eeq
where $\theta_{13}$ and $\theta_{23}$ are the corresponding CKM
mixing angles. The other generation (the first one) decouples,
and is protected by a residual $U(1)_Q$
symmetry~\cite{ourlong}.

As before, we wish to analyze the flavor violation induced by
$X_Q$ in a covariant form. The new contributions to $\Delta
t,b\,=\,1$ transitions are characterized by
\beq \label{inclusive_decay}
{\rm BR}(Q_3\to Q_{l})\propto \frac{4}{3}
{\left| X_Q \times \haud \right|^2}\, ,
\eeq
where $Q_l$ stands for light doublets. We stick to the same
definitions as in the two generation part,
Eq.~\eqref{definitions}. In the $[U(3)/U(2)]^2$ limit we can
covariantly identify four independent directions out of the
eight generators space: $\had$, $\hj$, $\hjd$ and an additional
one, $\hjq \equiv -2\had+\sqrt3\,\hj\times \hjd$. Since $\had$
and $\hjq$ are not orthogonal, we replace the former with
$\hadp\equiv \hj\times \hjd$ (and a similar expression for the
up sector).  Note that $\hjq$ corresponds to the conserved
$U(1)_Q$ generator, so it commutes with both $\ad$ and $\au$,
and takes the same form when interchanging $d\leftrightarrow u$
($\hj$ also remains the same in both up and down bases, as in
the two generations case). There are four additional
directions, collectively denoted as $\hat {\vec D}$, which
transform as a doublet of the CKM (2-3) rotation, and do not
mix with the other directions.

\section{Application~-- Third Generation Decays}
We next use measurements of down type FCNC and LHC projection
for top FCNC to derive a model independent bound on the
corresponding NP scale. We focus on the following operator
\beq \label{3g_operator}
O^h_{LL}= i \left[ \overline{Q}_i \gamma^\mu (\xqo)_{ij} Q_j
\right] \left[ H^\dagger \overleftrightarrow{D}_\mu H \right] +
\mathrm{h.c.} \, ,
\eeq
which contributes at tree level to both top and bottom
decays~\cite{Fox:2007in}\footnote{It is important to note that
a given new physics model might generate different
higher-dimensional operators via different types of processes.
Therefore $X_Q$ is in general different for each operator, so
we denote it specifically as $\xqo$ for the current case.}. We
adopt the weakest limits on the coefficient of this operator,
$C^h_{LL}$, derived in~\cite{Fox:2007in}:
\beq \label{exp_constraints}
\begin{split}
\mathrm{Br}&(B \to X_s\ell^+ \ell^-) \longrightarrow \left|
C^h_{LL} \right|_b < 0.018 \ltev^2 \, , \\ \mathrm{Br}&(t\to
(c,u)Z) \longrightarrow \left| C^h_{LL} \right|_t < 0.18
\ltev^2 \, ,
\end{split}
\eeq
where the latter is based on the prospect for the LHC bound in
the absence of signal, with 100\,fb$^{-1}$, and we define
$r_{tb} \equiv \left| C^h_{LL} \right|_t/\left| C^h_{LL}
\right|_b\,$.

The NP contribution can be decomposed in the covariant bases
\beq \label{xq_param}
\hspace*{-.25cm} X_Q \! = \! X'^{u,d} \haudp\hspace*{-.045cm}
+\hspace*{-.045cm} X^J \hj\hspace*{-.045cm} +\hspace*{-.045cm}
X^{J_{u,d}} \hjud\hspace*{-.045cm} +\hspace*{-.045cm} X^{J_Q}
\hjq\hspace*{-.045cm} +\hspace*{-.045cm} X^{\vec D} \hD .
\eeq
The weakest bound is obtained, for a fixed length $L\equiv
\left|X_Q\right|$, by finding a direction of $X_Q$ that
minimizes the contributions to $\left| C^h_{LL} \right|_t$ and
$\left| C^h_{LL} \right|_b$\,. It is clear, however, that
directions that contribute to first two generations flavor and
CPV at ${\cal O}\left(\lambda_{\rm C}\right)$ ($\lambda_{\rm
C}\sim 0.23$) are strongly constrained. Thus, the resulting
bounds would not correspond to the best alignment case. For
example,  when only $X^{J_Q}\neq0$, no third generation flavor
violation is induced. However, switching back on the light
quark masses, $X^{J_Q}$ (more precisely, a combination of
$X^{J_Q}$ and $X'^d$) does induce flavor violation between the
first two generations. At best it can be aligned with the down
mass basis, so that it contributes to $\Delta c=1$ transition
at ${\cal O}\left(\lambda_{\rm C}\right)$. The corresponding
bound is~\cite{ourlong}
\beq \label{3g2g_constraint}
L<0.59 \ltev^2; \ \  \Lambda_{\rm NP}>1.7 \, \mathrm{TeV} \, ,
\eeq
where the latter is for $L=1$. Similarly, it can be shown that
$X^{\vec D}$  yields $2\to1$ transitions when the contributions
to third generation decays are minimized. These cases,
therefore, do not represent the best alignment scenario.

\begin{figure}[hbt]
\centering
\includegraphics[width=0.45\textwidth]{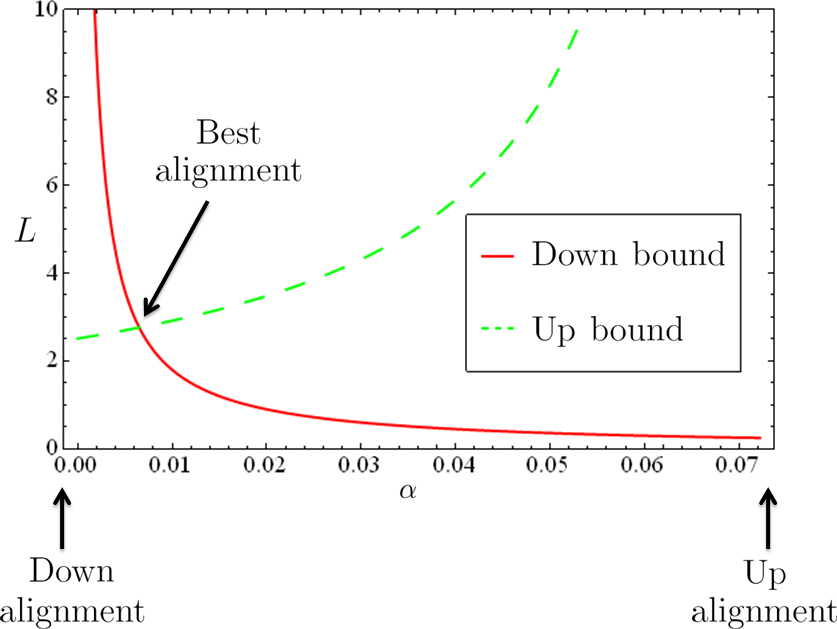}
\caption{Upper bounds on $L$ as a function of $\alpha$, coming from the measurements
of flavor violating decays of the bottom and the top quarks, assuming $\Lambda_{\rm NP}=1$\,TeV.}
\label{fig:3g_bounds}
\end{figure}

The induced flavor violation is then given by
\beq \label{explicit}
\frac{4}{3} \left| X_Q \times \haud \right|^2 =
\left(X^J\right)^2+ \left(X^{J_{u,d}}\right)^2  \,,
\eeq
and
\beq \label{explicit1}
X^{J_{u}}=\cos 2\theta \,X^{J_{d}}+ \sin 2\theta \,X'^{d}\,.
\eeq
From the above relations it is clear that $X^J$ contributes the
same to both rates, so it should be set to zero for optimal
alignment. Thus the best alignment is obtained by varying
$\alpha$, defined by
\beq
\tan\alpha \equiv X^{J_{d}}/X^d\,,
\eeq
where $X^d$ is the coefficient of $\had$, which is the
generator that does not produce flavor violation among the
first two generations to leading order (up to
$\mathcal{O}(\lambda_{\rm C}^5)$). We now consider two
possibilities: (i) complete alignment with the down sector;
(ii) the best alignment satisfying the bounds of
Eq.~\eqref{exp_constraints}, which gives the weakest
unavoidable limit. The bounds for these cases are
\beq \label{3g_bounds}
\mathrm{i)}&   \hspace*{-.35cm}\alpha=0 \, , \ L<2.5 \ltev^2; \
\Lambda_{\rm NP}\!>0.63 \,(7.9)\, \mathrm{TeV} \, ,
\\ \mathrm{ii)}& \alpha=\frac{\sqrt{3}\,\theta}{1+r_{tb}} \, , \ L<2.8 \ltev^2;
\ \Lambda_{\rm NP}\!>0.6 \, (7.6)\, \mathrm{TeV},\nonumber
\eeq
as shown in Fig.~\ref{fig:3g_bounds}, where in parentheses we
give the strong coupling bound, in which the coefficient of the
operators in Eqs.~\eqref{o1} and~\eqref{3g_operator} is assumed
to be $16 \pi^2$. Note that these are weaker than the bound in
Eq.~\eqref{3g2g_constraint}.

It is important to mention that the optimized form of $X_Q$
generates also $c \to u$ decay at higher order in $\lambda_{\rm
C}$, which might yield stronger constraints than the top decay.
In~\cite{ourlong} it is shown that the bound from the former is
actually much {\emph{weaker}} than the one from the top, as a
result of a $\lambda_{\rm C}^5$ suppression. Therefore, the LHC
is indeed projected to \emph{strengthen} the model independent
constraints.

\section{Third Generation $\Delta F=2$ Transitions}
Next we analyze $\Delta F=2$ processes, where for simplicity,
we only consider complete alignment with the down sector,
\beq
\xqt=L \had \,, \label{xq2}
\eeq
as the constraints from this sector are much stronger. This
generates in the up sector $D^0-\overline{D^0}$ mixing and top
flavor violation. Yet, there is no top meson, so we analyze
instead the process $uu \to tt$, which is most appropriate for
the LHC (and related to mixing by crossing symmetry). This
process was studied in the literature in the context of
different models (see
\textit{e.g.}~\cite{Larios:2003jq,Kraml:2005kb,Gao:2008vv} and
refs.~therein). It is observed through the dilepton mode, in
which two same-sign leptons are produced from the top quarks.
We emphasize that in this case the parton distribution
functions of the proton strongly break the approximate $U(2)$
symmetry of the first two generations. Thus, a useful bound is
obtained only from the operator involving up (and not charm)
quarks.

In order to estimate the prospect for the LHC bound on
same-sign tops production, we calculated the $uu \to tt$ cross
section using MadGraph/MadEvent~\cite{Alwall:2007st}, as a $t$
(or $u$) channel process mediated by a heavy vector boson,
matched onto the operator in Eq.~\eqref{o1}.  We then used the
fact that the cross section times the integrated luminosity
must be lower than 3 for a 95\% exclusion, in the absence of
signal~\cite{PDG}. Adding an assumption of 1\% signal
efficiency~\cite{Larios:2003jq}, after background reduction, we
have
\beq
z_1^{tt}<7.1 \times 10^{-3} \ltev ^2 \, ,
\eeq
for 100 fb$^{-1}$ at a center of mass energy of 14 TeV. The
experimental constraint from CPV in the D system
is~\cite{Gedalia:2009kh}
\beq
\mathrm{Im}(z_1^D)<1.1 \times 10^{-7} \ltev^2 \, .
\eeq

The contribution of $\xqt$ to these processes is calculated by
applying a simple CKM rotation, and then taking
$\mathrm{Im}\left[ \left(\xqt \right)^2_{12} \right]$ for CPV
in $D$ mixing and $\left| \left( \xqt \right)_{13} \right|^2$
for $uu \to tt$. The resulting bounds are
\beq
\begin{split}
L&<1.8 \ltev\,; \ \ \Lambda_{\rm NP}>0.57 \, (7.2)\,
\mathrm{TeV} \, , \\
L&<12 \ltev\,; \ \ \Lambda_{\rm NP}>0.08 \, (1.0)\,
\mathrm{TeV} \, ,
\end{split}
\eeq
for $D$ mixing and $uu \to tt$, respectively. Note that the
latter bound depends on the quartic root of the cross section
that was evaluated above, thus it is only mildly sensitive to
that calculation and to the efficiency assumption. Anyway, in
this case the existing bound is stronger than the one which
will be obtained at the LHC for top quarks, as opposed to
$\Delta F=1$ case considered above.

\section{Supersymmetry}
We now consider the application of our formalism to $\Delta
F=2$ transitions in supersymmetry (constraints from $\Delta
F=1$ processes are more involved, due to a richer operator
structure, and discussed in~\cite{ourlong}). We use the
approximation of quasi degenerate squark masses (see {\it
e.g.}~\cite{Ciuchini:1998ix}), and consider the leading order
in the expansion $\dmqot$, ($\dmqij$  is the mass-squared
difference between the $i$th and $j$th squarks), where the
level of degeneracy is much stronger~\cite{Blum:2009sk}. We
arrive at the following expression for the length of $X_Q$
\beq
L=\frac{\alpha_{s}}{18}\,\sqrt{\frac{g(x)}{2}} \,
\frac{\dmqoth}{\tilde m_Q^2}\,,
 \eeq
where $x=m_{\tilde{g}}^{2}/\tilde m_{Q}^{2}$, $m_{\tilde{g}}$
is the gluino mass and $g(x)$ is a known kinematic
function~\cite{Ciuchini:1998ix}. Taking for concreteness,
$\tilde m_Q=\left(2 m_{\tilde Q_1}+m_{\tilde Q_3}\right)/3$
(appropriate for models with only weak
degeneracy~\cite{Raz:2002zx}), $\tilde m_Q = 100\mbox{\,GeV}$
and $m_{\tilde{g}}\approx \tilde m_{{Q}}$, which implies
$g(1)=1$, we find
\beq
\frac{\left|m_{\tilde Q_1}^2- m_{\tilde Q_3}^2\right|}{\left(2
m_{\tilde Q_1}+m_{\tilde Q_3}\right)^2}
 <0.45\left(\frac{\tilde m_{Q}}{100\mbox{\,GeV}}\right)^2\,.
 \eeq
\section{Warped Extra Dimension}
Another example for a concrete model that is constrained by
measurements is the Randall-Sundrum (RS)
framework~\cite{Randall:1999ee}. When the fermions are allowed
to propagate in the bulk, their localization yields mass
hierarchies and mixing angles, thus addressing the flavor
puzzle. The $\Delta F=2$ process is induced at tree level by a
Kaluza-Klein (KK) gluon exchange. The $\Delta F=1$ operator in
Eq.~\eqref{3g_operator} is generated, among others, via mixing
between the SM Z and its KK excitations, which results in a
non-diagonal coupling in the mass basis~\cite{Agashe:2004cp,
Agashe:2006wa}. For simplicity, we only focus below on these
contributions, as the others are of the same
order~\cite{Agashe:2006wa}. For the $\Delta F=2$ case we have
\beq
m_{\rm KK}=\Lambda_{\rm NP} \, , \quad X_Q\cong
\frac{g_{s*}}{\sqrt{6}} \,\mathrm{diag}
(f^2_{Q^1},f^2_{Q^2},f^2_{Q^3}) \, ,
\eeq
before removing the trace, where $g_{s*}$ is the dimensionless
5D coupling of the gluon ($g_{s*} \approx 3$ at one
loop~\cite{Agashe:2008uz}) and the $f_{Q^i}$'s are the values
of the quark doublets on the IR brane. These are related to
each other through the CKM elements~-- $f_{Q^1,Q^2}/f_{Q^3}
\sim V_{ub},V_{cb}$. The resulting limit is
\beq
m_{\rm KK}>0.4 f^2_{Q^3} \; \mathrm{TeV} \, ,
\eeq
where $f_{Q^3}$ is typically in the range of 0.4-$\sqrt{2}$.
For the $\Delta F=1$ process we find
\beq
X_Q \cong g_{Z*}\, \delta g_Z\,
\mathrm{diag}(f^2_{Q^1},f^2_{Q^2},f^2_{Q^3}) \, ,
\eeq
where $g_{Z*}$ is the dimensionless 5D coupling of the Z to
left-handed up type quarks ($g_{Z*} \cong 1.2$ at one loop) and
$\delta g_Z \cong \log (M_{\mathrm{Pl}}/\mathrm{TeV})
\left(m_Z/m_{\rm KK} \right)^2$ describes the non-universal
coupling coming from mixing between the different Z states. The
bound that stems from this is
\beq
m_{\rm KK}>0.33 f^2_{Q^3} \; \mathrm{TeV} \, .
\eeq

\section{Conclusions}
We find that projected LHC bounds on $\Delta t=1$ processes
enable us to provide a new model independent constraint on the
strength of left-handed quarks flavor violation, even in the
presence of general flavor alignment mechanisms. The projected
bound on $\Delta t=2$ transitions from same sign tops
production at the LHC is also studied. In this case a
surprising result is that a stronger robust bound already
exists due to the experimental constraint on CP violation in
$D-\overline D$ mixing. We use our analysis to obtain new
limits on supersymmetric and warped extra dimension models of
alignment, which are rather weak (as a result of the weaker
experimental constraints, compared to the first two
generations~-- see {\it e.g.}\ in~\cite{Blum:2009sk}), but
replacing practically non-existing current bounds.

{\bf Acknowledgments } G.P. is supported by the Israel Science
Foundation (grant \#1087/09), EU-FP7 Marie Curie, IRG
fellowship and the Peter \& Patricia Gruber Award.

\end{document}